# Prediction of Apophis Asteroid Flyby Optimal Trajectories and Data Fusion of Earth-Apophis Mission Launch Windows using Deep Neural Networks

*Manuel Ntumba [(1)], Saurabh Gore [(2)], Jean-Baptiste Awanyo [(3)]*

[(1)(2)(3)] *Division of Space Applications, Tod'Aers, Lomé, Region Maritime, Togo.*

[(1)] *Space Generation Advisory Council, Vienna, Austria.*

[(2)] *Moscow Aviation Institute, Moscow, Russia.*

[(3)] *Univerité Sultan Moulay Slimane, Béni Mellal, Morocco.*

manuel.ntumba@spacegeneration.org

sdgore@mai.education

## ABSTRACT

In recent years, understanding asteroids has shifted from light worlds to geological worlds by exploring modern spacecraft and advanced radar and telescopic surveys. However, flyby in 2029 will be an opportunity to conduct an internal geophysical study and test the current hypothesis on the effects of tidal forces on asteroids. The Earth-Apophis mission is driven by additional factors and scientific goals beyond the unique opportunity for natural experimentation. However, the internal geophysical structures remain largely unknown. Understanding the strength and internal integrity of asteroids is not just a matter of scientific curiosity. It is a practical imperative to advance knowledge for planetary defense against the possibility of an asteroid impact. This paper presents a conceptual robotics system required for efficiency at every stage from entry to post-landing and for asteroid monitoring. In short, asteroid surveillance missions are futuristic frontiers, with the potential for technological growth that could revolutionize space exploration. Advanced space technologies and robotic systems are needed to minimize risk and prepare these technologies for future missions. A neural network model is implemented to track and predict asteroids' orbits. Advanced algorithms are also needed to numerically predict orbital events to minimize errors.

## 1. INTRODUCTION

Apophis is a near-Earth asteroid with 370 meters of diameter, which initial observations indicated a probability of up to 2.7% that it would hit Earth in 2029. Additional observations provided improved predictions that ruled out the possibility of an impact on Earth in 2029. However, until 2006, there was still the possibility that Apophis would pass through a gravitational keyhole, a small region no longer about 800 meters in diameter [1] [2] which would establish a future impact exactly seven years later in 2036. It was considered at level 1 on the Turin Impact Risk Scale until August 2006, when the probability of Apophis going through the keyhole was judged to be very low, and the Apophis rating on the Turin scale was lowered to zero. In 2008, it was determined that the keyhole was less than 1 km wide. Apophis set the record for the highest score on the Turin Scale during the shortest period of greatest concern, reaching Level 4 on December 27, 2004. [3] Understanding asteroids and their risk of impact are some of the great responsibilities and challenges of our time. Natures are cooperating by offering once for a thousand years the opportunity to study the result of an extremely close passage of an unprecedented 350-meter asteroid, called Apophis, in 2029. The close encounter of Apophis will take place inside the Earth's geosynchronous ring of satellites at a distance less than a tenth of the lunar distance. While previous spacecraft missions have studied asteroids, none have ever had the opportunity to study live results of planetary tidal forces on their shapes, rotational states, surface geology, and internal structure. These physical parameters, and their changing response to induced stresses, represent an incredible opportunity to gain vital knowledge to deal with the eventuality of a known asteroid on an actual impact trajectory. Earth-Apophis Mission can monitor its surface and interior structure before, during, and after its near-Earth encounter in 2029.

## 2. MISSION OVERVIEW

For Earth-Apophis, the launch period is the collection of days, and the launch window is the period during which the mission must be launched to reach its target. [14] [15] If the rocket is not launched in a given window, it must wait for the window the next day of the period. [16] The launch periods and launch windows depend on both the rocket's capacity and the orbit to which it is heading. [17] [18] The launch window is defined by the first launch point and the final launch point, capable of launching every second in the launch window. [19] The spacecraft intends to surrender with the Apophis asteroid already in orbit; the launch must be carefully timed to occur as the target vehicle's orbital plane intersects the launch site. [20]

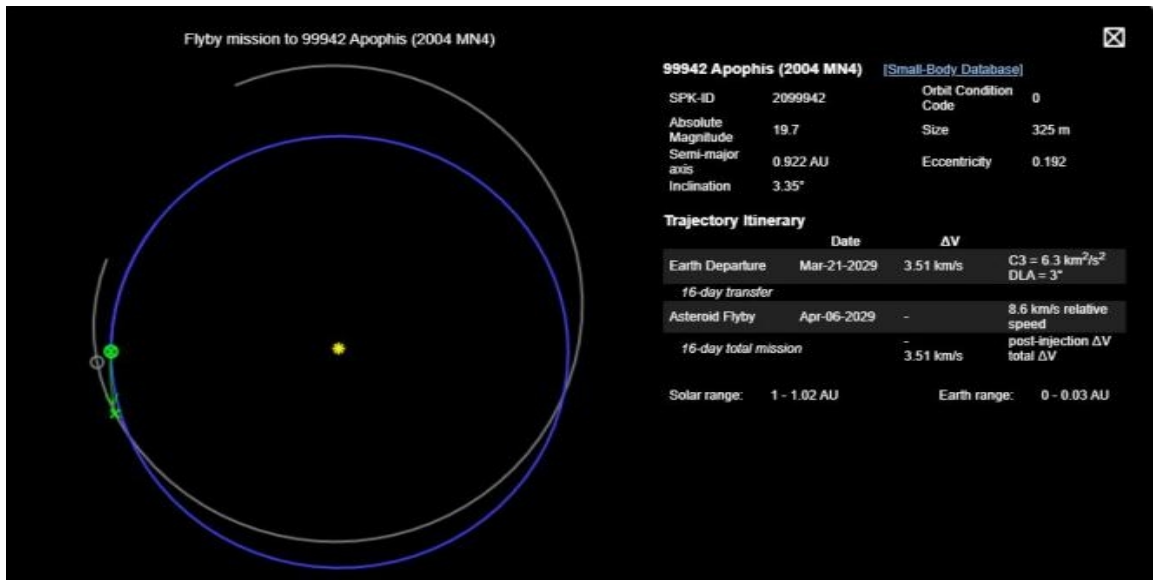

**FIGURE 1: Apophis Asteroid Flyby Optimal Trajectories.** The closest known approach to Apophis occurs on April 13, 2029, 21:46 UT, when Apophis will pass the Earth closer than the geosynchronous communication satellites but will not come close to 31,600 kilometers above the surface of Earth. [6] Using the March 2021 orbital solution, which includes the Yarkovsky effect, the 3-sigma region of uncertainty in the 2029 approach distance is approximately ± 2 km. [4] The distance, the width of a hair in astronomical terms, is five times the Earth's radius and ten times closer than the Moon. Apophis asteroid will become as bright as the magnitude visible from darker rural and suburban areas, visible with binoculars from most places. [7][5][8]

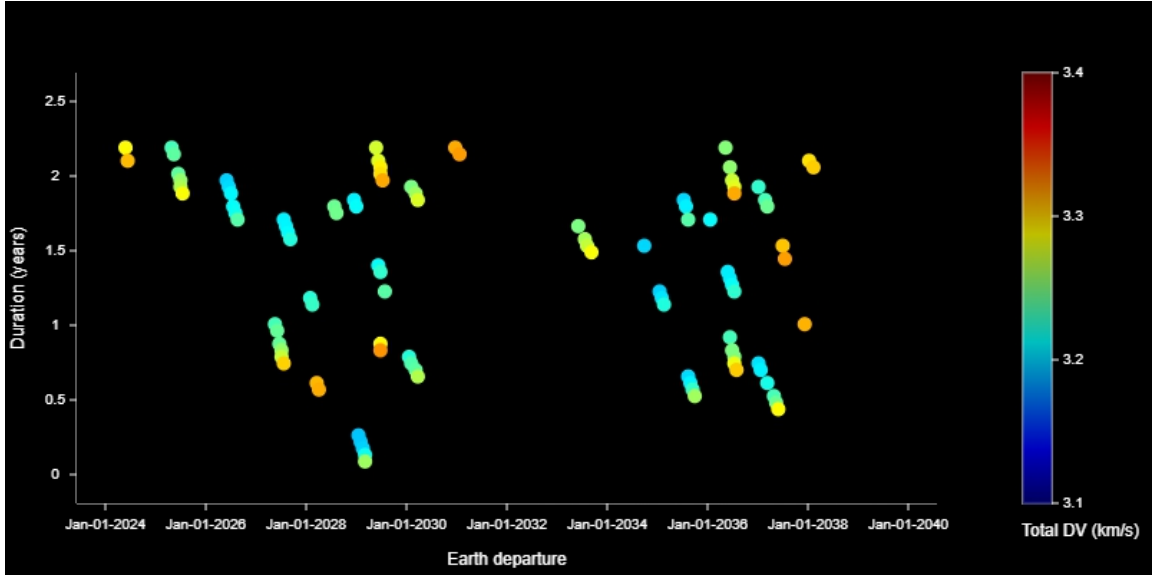

**FIGURE 2: Earth-Apophis Mission Launch Windows.** The close-up approach will be visible from Europe, Africa, and West Asia. Earth-Apophis Mission approaches between an Aten-class orbit with 0.92 AU semi-major axis and an Apollo-class orbit with 1.1 AU a semi-major axis. [9] Perihelion will decrease from 0.746 AU to 0.894 AU, and aphelion will decrease from 1.099 AU to 1.31 AU. In 2036, Apophis will approach Earth a third the distance from the Sun in March and December. [4] Using the 2021 orbit solution, Earth's approach in 2036 will not be closer to 0.3095 AU but more likely around 0.3100 AU. [10]

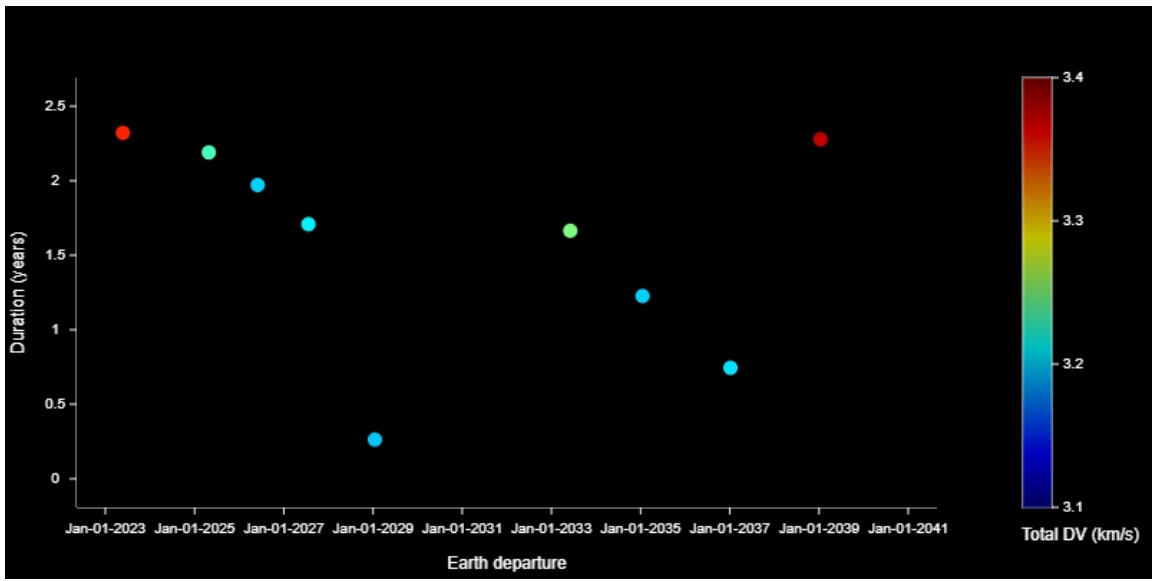

**FIGURE 3: Earth-Apophis Mission Optimal Trajectories.** The trajectory optimization in the Earth-Apophis Mission includes designing trajectories that minimize or maximizes some measure of performance according to a set of constraints. [11] It is often used for systems where the complete closed-loop solution's computation is unnecessary, impractical, or impossible. The trajectories can be optimized by including the inverse of Lipschitz's constant [12], then it can be used to generate a closed-loop solution. If only the first step of the path is performed for an infinite horizon problem, it is called Model Predictive Control (MPC). Many of the original

trajectory optimization applications were in the aerospace industry, calculating the launch trajectories of rockets and missiles. More recently, path optimization has also been used in various industrial processes and robotics applications. [13]

### 3. METHODS AND RESULTS

The DNN models implement the process of integrating multiple data sources to produce more advanced optimizations. The data fusion processes are often categorized as low, medium, or high, depending on the processing stage at which the fusion takes place. [22] Low-level data fusion combines multiple raw data sources to produce new raw data. The data should be more informative and synthetic than the original entries. [23] Data fusion is often necessary to combine various data sets into a dataset that includes all input datasets' data points and time steps. The merged dataset is different from a simply combined superset in that the points in the merged dataset contain attributes and metadata that may not have been included for those points in the dataset. For Earth-Apophis, the launch window occurs when the launch site location is aligned with the plane of the orbit of Apophis. Launching at any other time would require an orbital plane change maneuver that would require a large amount of thruster. For launches above low earth orbit (LEO), the actual launch time may be somewhat flexible if a parking orbit is used, as the inclination and time that the spacecraft initially spends over The parking orbit can be changed. [14] Reaching the desired orbit requires the accurate Right Ascension of the Ascending Node (RAAN). RAAN is defined by varying a launch time, waiting for the earth to spin until it is in the correct position. For missions with very specific orbits, the launch window can be a single instant in time, called the instant launch window. The trajectories are programmed in a launcher before launching. The launcher will have a target, and the guidance system will alter the steering controls to arrive at the end state. At least variables such as apogee, perigee, inclination must be left free to modify the others' values; otherwise, the dynamics would be too constrained. An instant launch window allows the RAAN to be the uncontrolled variable. While some spacecraft, such as the upper stage, can steer and adjust their RAAN after launch [21], choosing an instant launch window allows the RAAN to be predetermined for the spacecraft guidance system.

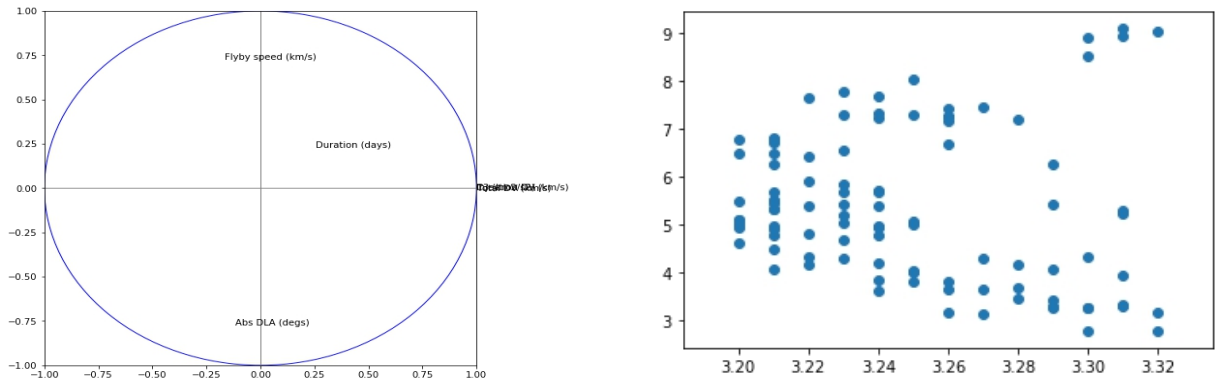

**FIGURE 4. Correlation using Deep Neural Networks.** The correlation shows that the Injection DV (km/s) and the Total DV (km/s) have almost the same value. Also, the Flyby speed (km/s) and the Abs DLA (degs) are negatively correlated. Hence the increase of one leads to the decrease of the other. Supervised learning of neural networks uses a set of corresponding inputs to produce the desired output for each input. The cost function is related to the elimination of incorrect deductions. [24] A commonly used cost is the root mean square error, which attempts to minimize the root mean square error between the network and the desired

output. Based on this plot we can not an application that connects Total DV (km/s) and Duration (days) for a single value of x there are several values of y which correspond so we can conclude that there are other factors which influence the variation of y therefore a model based on x to predict the value of y would not be efficient.

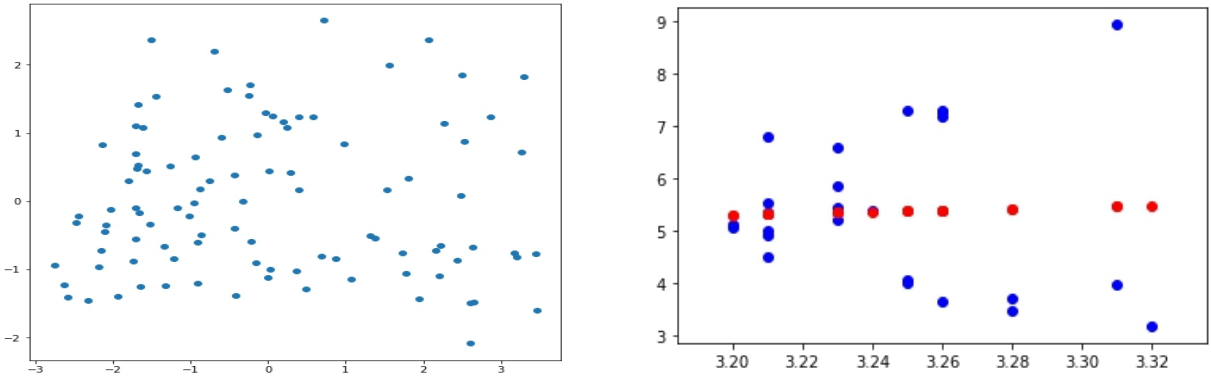

**FIGURE 5. Prediction and Data Fusion using Deep Neural Networks.** From that graph we can see the huge gap between the predicted values and the actual values*.* The DNN models suitable for supervised learning are pattern recognition or classification and function regression or approximation. Supervised learning also applies to sequential data, which provides continuous feedback on the quality of the solutions obtained. In unsupervised learning of neural networks, the input data is provided with the cost function, a data function, and the network's output. The cost function depends on the DNN model and the prior assumptions or the model's implicit properties, parameters, and the observed variables. DNN models that fall under the unsupervised learning paradigm are usually estimation problems; applications include grouping, estimation of statistical distributions, compression, and filtering.

## 4. CONCLUSIONS

The Earth-Apophis mission will take advantage of the incredible opportunity that nature offers to study the impact of tidal interactions on potentially dangerous asteroids. The mission will launch in March 21$^{st}$ 2029 and arrive at Apophis in April 6$^{th}$ 2029. The Earth-Apophis mission shows that a scientifically robust mission is well within the range of flight equipment and the proven high heritage value launch capability currently available. The scientific results can directly inform future studies on the mitigation of asteroids' impact, including long-term monitoring correlating the measured thermal emission and the corresponding Yarkovsky drift. Space robotics is a relevant tool for space exploration. It helps to overcome human borders in space. The past decade has seen tremendous advancements in space robotics and its applications for space missions. We propose a neural network model implemented for the Earth-Apophis asteroid mission. By including a thermal instrument and continuing to orbit Apophis after encountering Earth, Earth-Apophis will monitor and decode the coupling of rotation and thermal cycle resulting from Yarkovsky's drift. The direct correlation of thermal properties with the resulting Yarkovsky drift is important for future Apophis orbital predictions and for improving the general understanding of asteroid dynamics. The Earth-Apophis orbiter will also map the global geology and composition of Apophis and study its interior structure, improving knowledge of medium-sized asteroids 100 meters in diameter. Studies of asteroids by spacecraft can provide insight into the geological and dynamic history of the objects they study, improve our understanding of these individual objects, and have important implications for understanding the solar system's formation.